# WATERMARKING USING DECIMAL SEQUENCES

Navneet Mandhani and Subhash Kak

ADDRESS: Department of Electrical and Computer Engineering, Louisiana State University, Baton Rouge, LA 70803, USA.

**ABSTRACT:** This paper introduces the use of decimal sequences in a code division multiple access (CDMA) based watermarking system to hide information for authentication in black and white images. Matlab version 6.5 was used to implement the algorithms discussed in this paper. The advantage of using d-sequences over PN sequences is that one can choose from a variety of prime numbers which provides a more flexible system.

## 1. INTRODUCTION
Digital watermarking includes a number of techniques that are used to imperceptibly convey information by embedding it into cover data [1]. In the problem of establishing the identity of the owner of an object, identity may be established by either printing the name or logo on it. Unlike printed watermarks, digital watermarking is a technique where bits of information are embedded in such a way that they are completely invisible. The problem with the traditional way of printing logos or names is that they may be easily tampered or duplicated. In digital watermarking, the actual bits are scattered in the image in such a way that they cannot be identified and show resilience against attempts to remove the hidden data [1].

Decimal sequences are obtained when a number is represented in a decimal form in a base *r*. For a certain class of decimal sequences of *1/q*, *q* prime, the digits spaced half a period apart add up to *r*-1, where *r* is the base in which the sequence is expressed. Decimal sequences have good autocorrelation and cross correlation properties and they may be used in applications involving PN sequences [2] [3]. The following sections describe a few properties of decimal sequences, their generation using feedback shift registers that allow carry and their application to watermarking using black and white images.

## 2. GENERATION OF DECIMAL SEQUENCES
Decimal sequences can be generated by using feedback shift registers that allow carry. The simplest way to generate a binary d–sequence is by using the equation $a^i = (2^i \bmod q) \bmod 2$. The hardware used for their generation is similar to that used for PN–sequences, and it is based on an algorithm called the Tirtha algorithm [4], which may be used whenever the prime number *q* is given in terms of the radix *r* as $q = tr - 1$, where *t* is an integer.

Theorem: Consider that $1/(tr - 1)$ defines the d–sequence $a_1a_2a_3....a_k$, where $r$ is the radix or the base. Consider another sequence $u_1u_2u_3....u_k$, where, for all $i$, $u_i < t$, then
$$ru_i + a_i = u_{i+1} + ta_{i+1}$$

Proof: Since the sequences repeats itself $a_k = 1$ and $u_k = 0$. The remainder in the long division of 1 by $(tr - 1)$ is therefore $t$. The quotient $a_{k-1}$ is given by
$$a_{k-1}(tr - 1) + t = m_{i-1} r$$
This makes $a_{k-1} = t$, extending the argument the $a$ and $u$ sequences, when written in inverse as
$$u_k u_{k-1}.....1$$
$$a_k a_{k-1}......0$$
equal
$$0\ 0\ .....1$$
$$1\ t\ [t^2] mod\ r......0$$

Example:
Consider $\{1/19\} = 1/(2 \times 10 - 1))$ you mean $1/(2 \times 10 - 1))$ in the base 10. The inverse sequence is then given as it would be helpful if you would step through one or two of these
$$0\ 0\ 0\ 0\ 1\ 1\ 0\ 1\ 0\ 1\ 1\ 1\ 1\ 0\ 0\ 1\ 0\ 1$$
$$1\ 2\ 4\ 8\ 6\ 3\ 7\ 4\ 9\ 8\ 7\ 5\ 1\ 3\ 6\ 2\ 5\ 0$$

In the lower row, we multiply by 2 as we move to the right and add to the result the corresponding digit in the first row, but placing only the "units" in the lower row and the "tenths" in the top row. Thus, 8 in the fourth column, lower row, becomes 6 at the next location, with a 1 in the top row. The next entry would be $6 \times 2 + 1 = 13$, which means 3 in the bottom row and 1 in the top row.
Or, the d–sequence for $\{1/19\}$ is given by writing the digits in the lower row in the reverse order:
$$0\ 5\ 2\ 6\ 3\ 1\ 5\ 7\ 8\ 9\ 4\ 7\ 3\ 6\ 8\ 4\ 2\ 1$$
The circuit for the generation of d–sequences $1/(tr - 1)$ is given in Figure 1.
It consists of n stages of shift registers. The c's represent carries that are added to the immediately preceding stages. When the carry is generated by the extreme left stage, it is introduced into this stage at the very next clock instant. As explained before, the sequence generated will be in the reverse order. The same principle can be used to generate binary d–sequence. The number of stages needed for the generation of binary d–sequence $1/q$ is about $log_2 q$. The algorithm also works for the non binary sequences of the type $1/(tr - 1)$ when the given fraction is multiplied by an appropriate integer so that the standard form can be used.

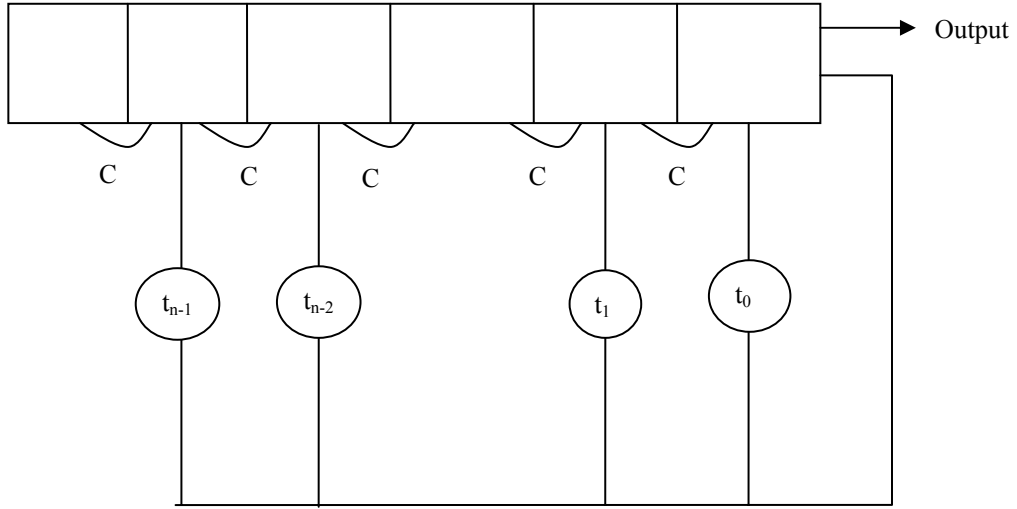

Fig 1.  Generation of d-sequences

**3. EMBEDDING AND DECODING ALGORITHM**
Since, decimal sequences have zero cross correlation for some prime numbers [2], one would obtain superior performance if different d-sequences are used in the watermark. But, if the same d-sequence is used, the autocorrelation can be as high as 33% during certain shifts of the sequence. However, since many other shifts have zero autocorrelation, we can selectively use these shifts to produce excellent results [5]. The use of decimals sequences also gives the flexibility of trying out various prime numbers until we get satisfactory embedding and recovery of the hidden information.

**Embedding**

A decimal sequence is generated in Matlab using the function
$$\text{dseq} = [r^i \bmod q] \bmod r$$
Where, $r$ is the radix or the base and $q$ is the prime number.
The d-sequence generated is used for embedding the data in the cover image. This helps us exploit the auto-correlation properties of the d-sequences. The addition of the d-sequences to the cover image is done according to the equation:

$$I_w(x, y) = I(x, y) + k \times W(x, y)$$

Where,
$I_w(x, y)$ denotes the watermarked image.
$I(x, y)$ denotes the actual cover image.
$W(x, y)$ denotes a decimal sequence noise pattern that is added to the image.
$K$ denotes the gain factor.
To show how the above method works consider watermark image ($a(x,y)$) as the information bearing data signal and d-sequence ($b(x,y)$) as the spreading signal. The desired modulation is achieved by applying both the watermark image and the d-sequence to a addition(?) modulator. The resultant signal W(x,y) is a decimal sequence noise pattern that is added to the cover image I(x,y) to produce the resultant watermarked image $I_w(x, y)$.

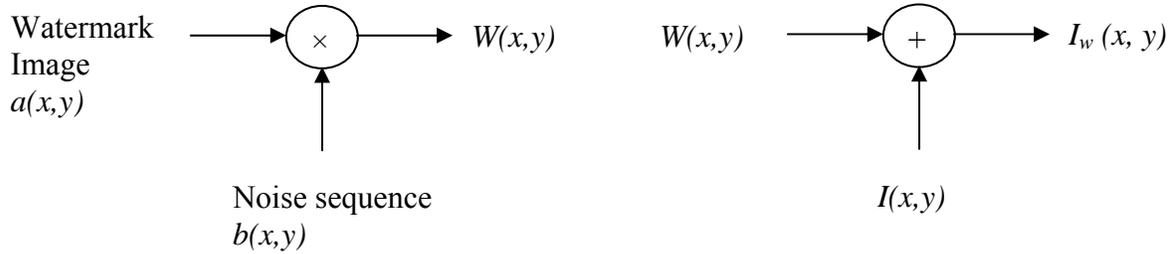

Fig 2 Embedding process

Hence $I_w(x, y) = K \times W(x,y) + I(x,y)$
$= a(x,y) \times b(x,y) + I(x,y)$

**Decoding**
To recover the original watermark $a(x,y)$, the watermarked image $I_w(x, y)$ is multiplied at the receiver again with a decimal sequence which is an exact replica of that used for embedding the data.

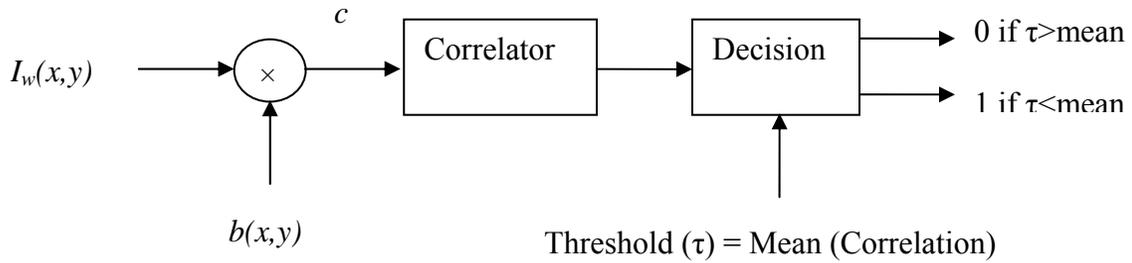

Fig 3 Recovery process

The multiplier output $C$ is given by

$$C = I_w(x,y) \times b(x,y)$$
$$= (a(x,y) \times b(x,y) + I(x,y)) \times b(x,y)$$
$$= a(x,y) \times b^2(x,y) + I(x,y) \times b(x,y)$$

The above equation shows that the watermark image $a(x,y)$ is multiplied twice with the noise signal $b(x,y)$, whereas the unwanted or the cover image $I(x,y)$ is multiplied only once with the noise signal. So $b^2(x,y)$ becomes 1 and the product $I(x,y)*b(x,y)$ is the unwanted noise signal that can be filtered out during the process of correlation by setting the threshold as mean of correlation. Hence, at the receiver we recover the watermark image $a(x,y)$ [6].

There is a trade off between the robustness and the quality of the image as the gain K is increased. But, the decimal sequences gives us an option of experimenting with various prime numbers, keeping the gain constant, until we observe a satisfactory result for both encryption of data and its retrieval. For retrieval of the encrypted message the decimal sequences are generated again and then correlated with the watermarked image. The d-sequences may be added to the cover image by either a circular shift or a random shift.

Sample results for random shifts as defined above are shown below for the image and watermark of Figure 4, where the second figure represents the watermark.

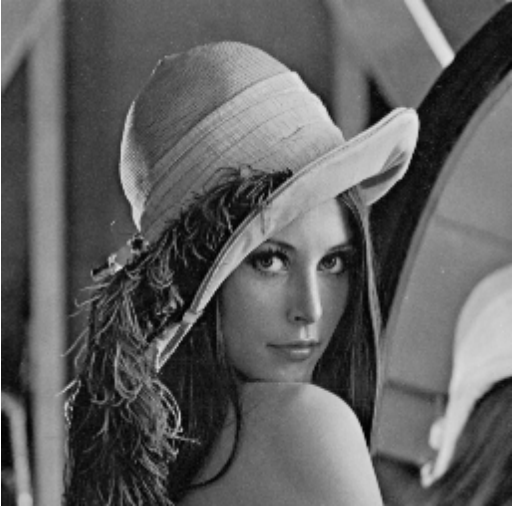

Fig 4
256 × 256 (8 bit)
Gray scale Lena image

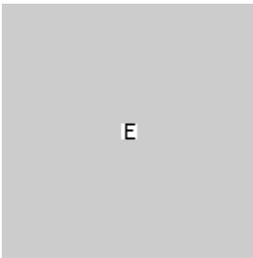

8 × 8 Watermark object
Monochrome Image

Figures 5 and 6 give the image with the correspondingly retrieved watermark for q=283

and q=167. Equally good results are obtained if the period of the sequence is long.

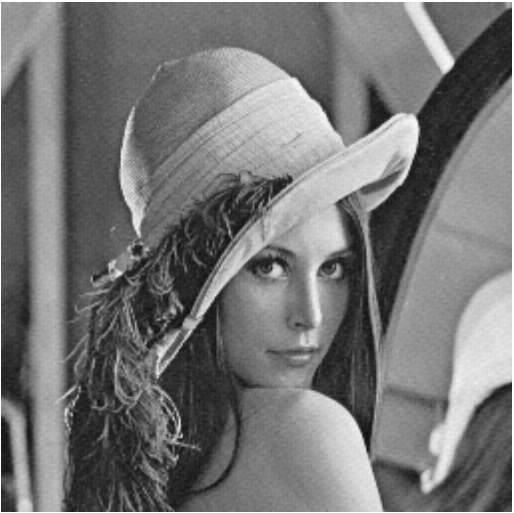

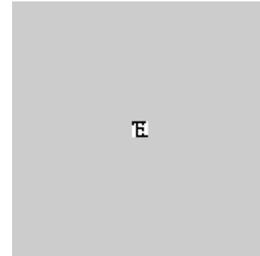

Fig 5
Embedding output, q = 283

Decoding output,
Period = 94

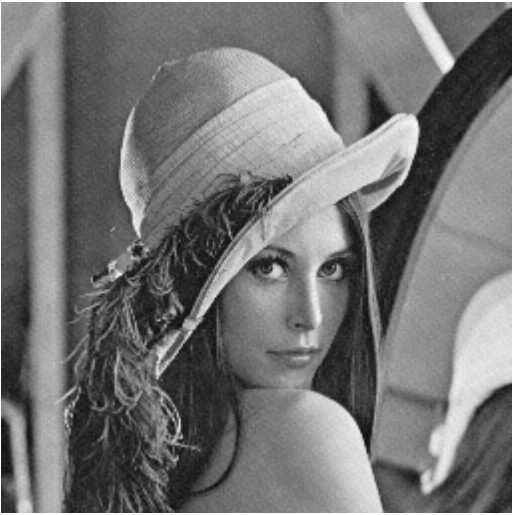

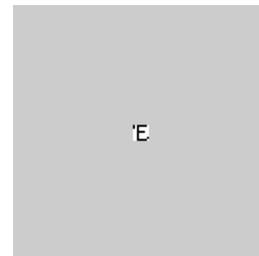

Fig 6 Embedding output, q = 167

Decoding output,
Period = 84

## 4. PERFORMANCE ANALYSIS

To analyse the performance of decimal sequences for watermarking we stress that for certain choices of q and period, the results are not satisfactory. Our experiments have shown that the increase in the number of noise pixels is due to the decrease in the period of decimal sequence. From Fig 7 it is evident that for autocorrelation values lying between the standard deviation range good recoveries are possible while for autocorrelation values lying very close to the mean the best possible recovery is obtained.

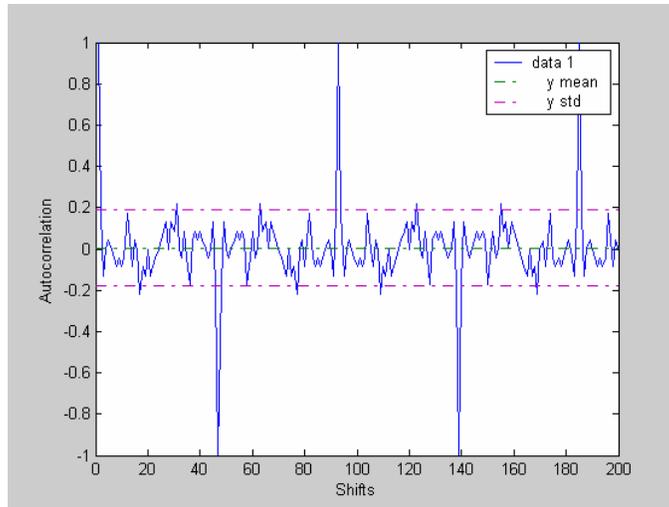

Fig 7 Autocorrelation for *q*=277

For the above case the best possible recovery is possible with a shift of 74 as shown in Fig 8. As seen from the graph in Fig 7 the autocorrelation value for a shift of 74 is closest to zero. But if the sequence is rotated by a shift of 120 the retrieval of the hidden watermark is unrecognizable (as shown in Fig 9) because of the high value of autocorrelation at shift 120. The best possible recovery for a near to zero autocorrelation does not necessarily mean a perfect recovery. A perfect recovery is also dependent on the period of the sequence.

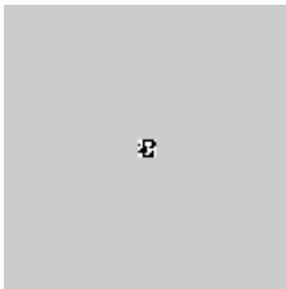      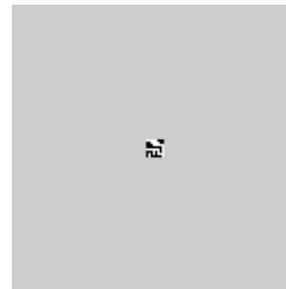

   Fig 8 Rotation 74                                                                                    Fig 9 Rotation 120

The noise associated with the retrieval of watermark is due to the fact that some pixels have correlation values very close to the mean correlation, which is set as the threshold. The retrieval algorithm takes it as a black pixel which results as a noise pixel in the recovery. On the other hand, a perfect recovery has exactly the number of black pixels in the watermark above the threshold.

Our experiments indicate that as the period of the generated decimal sequence reduces, the noise associated with the recovery increases. This is in turn associated with reduced randomness of the sequence with reduction in period. This causes a possible increase in the correlation value for pixels and thus causes more noise at the output.

## 5. CONCLUSIONS

This paper presents techniques of watermarking using decimal sequences. Since these sequences have zero autocorrelation for certain shifts, this could be useful in the recovery of watermarks by using spread spectrum techniques. Use of these sequences over PN sequences provides the following advantages:

- Hardware complexity can be reduced by not incorporating the high pass filter which might be needed in case of PN sequences when the recovery is not optimal.
- Performance of the d-sequence watermarking can be improved by using a variety of prime numbers with varied periods and particular shifts that provide close to zero autocorrelations.
- Decimal sequences exhibit zero cross correlation for some prime numbers and near to zero cross correlation for others, which would be useful if different d-sequences are used in the watermark.
- There is a trade off between robustness and perceptibility of the watermarked image as the gain K is increased for both d-sequences and the PN sequences but, for d-sequences we can use a different prime number with varied periods rather than accepting reduced performance.

This paper is limited to watermarking of still black and white images. Further research may be done for developing watermarking techniques for audio and video images, network packets, software and circuitry.